\documentclass[global,twocolumn]{svjour}

\usepackage{graphicx}
\journalname{Applied Physics A}
\begin{document}

\title{Field Induced Partial Order in the Spin-Ice Dysprosium Titanate.}
\author{T. Fennell\inst{1} \and O.A. Petrenko\inst{2}\thanks{\emph{Present Address: Department of Physics, University of Warwick, Coventry, CV4 7AL.}} 
\and  G. Balakrishnan\inst{3} \and
S.T. Bramwell\inst{4}\and J.D.M. Champion\inst{4} \and B. F{\aa}k\inst{2}
\and
M.J. Harris\inst{2} \and D. McK. Paul\inst{3}}

\offprints{T. Fennell, tom@ri.ac.uk} 

\institute{Davy-Faraday Research Laboratory, Royal
Institution of Great Britain, 21 Albemarle Street, London W1S 4BS, \and
ISIS Facility, Rutherford Appleton Laboratory, Chilton, Didcot, OX11 0QX, \and
Department Of Physics, University of
Warwick, Coventry, CV4 7AL, \and
Department of Chemistry, University College, 20 Gordon Street, London,
W1C 0HJ.}
\date{Received:16$^{th}$ July 2001  / Revised version: 16$^{th}$ July
2001}
%
\maketitle
\begin{abstract}
Neutron scattering has been used to investigate the 
magnetic correlations in the the spin ice
material dysprosium titanate, Dy$_2$Ti$_2$O$_7$. 
An isotopically enriched sample was used to minimise
neutron absorption.  In zero field no
magnetic order was observed down to 50 mK but the magnetic 
diffuse scattering was in
qualitative agreement with that expected for
the disordered low temperature state of dipolar spin ice. 
Application of a field of $\approx 0.8$T in the
$[100]$ direction led to long range order.  With the field  
applied in the $[1\bar{1}0]$ direction a coexistence of long range
ferromagnetic and short range antiferromagnetic order was observed. 
This is attributed to the pinning of only half the 
spins by the field.  The hysteresis loops in both field orientations 
displayed unusual steps and plateaus. 

75.25.+z,75.40.Gb,75.50.Lk
 \end{abstract}
\section{Introduction}
\label{intro}

Ho$_2$Ti$_2$O$_7$ and Dy$_2$Ti$_2$O$_7$ 
are regarded as good examples of
spin-ice materials~\cite{prl1,newprl,art}. 
The magnetic rare earth ions
occupy a pyrochlore lattice, a cubic array of corner-linked tetrahedra. 
In the near neighbour spin ice model the lattice is occupied by 
$\langle 111 \rangle$ Ising spins that are coupled ferromagnetically 
to nearest neighbours~\cite{jpcm}.  The result is a disordered, 
macroscopically degenerate, ground state that is determined only by the rule
that two spins must point into and two out of each tetrahedron. 
This ``two in, two out'' rule is analogous to the ``ice rules'' 
that control
the proton arrangement in
water ice, and so the near neighbour 
spin ice model maps exactly onto 
Pauling's model of the proton disorder in water ice \cite{jpcm,pauling}. 
The near neighbour spin ice model qualitatively describes much of the
behaviour of the real materials, but for an accurate description the dipolar
spin ice model has been developed \cite{byron}. 
In this model,
the near neighbour ferromagnetic coupling 
between $\langle 111 \rangle$ spins is dipolar in origin
and it has been shown that the long range part of the dipolar 
coupling maintains a spin-ice like disordered ground state 
even in the presence of an antiferromagnetic near neighbour 
super-exchange \cite{byron}. 
The dipolar spin ice model has been 
shown to give an accurate description of the zero field 
magnetic neutron scattering of Ho$_2$Ti$_2$O$_7$~\cite{newprl} in the spin ice
regime ($T < 2$ K) as well as
the specific heat of Ho$_2$Ti$_2$O$_7$ and Dy$_2$Ti$_2$O$_7$ throughout the
low temperature range (50 mK - 20 K) \cite{newprl,byron}.  The dipolar 
coupling in Ho$_2$Ti$_2$O$_7$ and Dy$_2$Ti$_2$O$_7$ is 
of similar magnitude 
($D_{nn}\approx2.35$K for both Dy and Ho) while 
the antiferromagnetic exchange is weak in both cases  
(-1.2K for Dy and -0.52K for Ho)~\cite{newprl,byron}. 
The main difference between these two materials appears 
to arise from the nuclear spins. Ho$_2$Ti$_2$O$_7$ has a single 
isotope ($^{165}$Ho) and strong hyperfine coupling that complicates the spin 
ice freezing process below $\approx0.8$ K \cite{newprl}. 
Dy$_2$Ti$_2$O$_7$, on the other hand, has much weaker 
hyperfine coupling that does not affect the dominant spin ice behaviour in 
the temperature range of interest. It 
might therefore be considered the simpler spin ice material.  
However, the natural mixture of Dy isotopes leads to 
significant neutron absorption (994 barn for 2200ms$^{-1}$ neutrons).
Thus, although neutron scattering 
should be the ideal microscopic probe of the spin correlations and 
dynamics of this system, 
the most significant experimental observations so far are 
bulk measurements~\cite{art}. To minimise the absorption problem we have 
prepared a single crystal of
$^{162}$Dy$_2$Ti$_2$O$_7$. 
In this paper we present the initial results of our
neutron scattering study of this sample, in both zero and applied 
magnetic field. 

\section{Experimental}
\label{exp}
The absorption cross section of natural Dysprosium is $\sigma_a = 994$ barn, 
the large value being 
due mainly to the presence of $^{164}$Dy.
Isotopically enriched Dy$_2$O$_3$ with composition
$^{156}$Dy$ < 0.01 \%$,
$^{158}$Dy$ < 0.01 \%$,$^{160}$Dy$ = 0.02 \%$,\\$^{161}$Dy$ = 0.47 \%$,
$^{162}$Dy$=  96.8 \%$,$^{163}$Dy$ = 2.21 \%$ and $^{164}$Dy$\\ = 0.5 \%$ was 
prepared by Goss Scientific Instruments Ltd.  
Enrichment in  $^{162}$Dy was considered optimum as it has a high abundance 
and the dominant impurity is $^{163}$Dy which has a slightly lower absorption cross section.  Enrichment in $^{163}$Dy would have lead to $^{164}$Dy as the dominant impurity.  Through enrichment, the absorption cross section was reduced by a factor of five to $208 \pm 13$ barn.

A single crystal of Dy$_2$Ti$_2$O$_7$ was prepared from the isotopically
enriched Dy$_2$O$_3$ and TiO$_2$ by the floating zone technique \cite{don}.
Neutron scattering was
carried out at the ISIS facility on the indirect geometry spectrometer PRISMA. It was configured in the 
diffraction mode so that 
sixteen $^3$He tube detectors are used simultaneously. 
Rotation of the crystal allows a rapid mapping of a
large section of reciprocal space, making PRISMA ideal for observing
magnetic diffuse scattering.

In the first experiment the crystal was cooled by a $^3$He sorption
refrigerator for measurements in zero applied magnetic field. It was 
aligned with $[1\bar{1}0]$ vertical, such
that the scattering plane contained $(hhl)$ wavevectors.
In the second experiment an Oxford instruments 7T 
vertical field cryomagnet, with dilution refrigerator
insert, was used. For the applied field measurements
two field orientations were studied: $[1\bar{1}0]$ as above and $[100]$
corresponding to a $(0kl)$ scattering plane. 
The data were normalised to monitor counts and vanadium to remove the 
characteristic flux profile of the spallation source.  Since the sample 
is still significantly absorbing a correction for this was also applied.

\section{Results}
\label{res}

With $[100]$ vertical, a map of reciprocal space was made
at the base temperature ($\approx 70$ mK) in zero field. The diffuse 
scattering maxima observed agree with those predicted in the $(hhl)$ 
scattering plane for the dipolar spin ice model.  For example there is a 
diffuse feature at $3,0,0$ in $(hk0)$ and $0,0,3$ in $(hhl)$~\cite{newprl}. 
On application of a field the diffuse scattering
disappeared and was replaced by magnetic Bragg peaks at the $Q = 0$ 
positions.  A field of $\approx0.7$T was sufficient to saturate these peaks.
  
This behaviour is readily explained as the applied field breaks the 
degeneracy of the six ``two in, two out'' spin configurations of the
elementary tetrahedron. The pyrochlore lattice can be described as a face
centred cubic lattice with a tetrahedral basis, and the 
degeneracy breaking means that every tetrahedron
adopts the same ``two in, two out'' state
with a net moment in the direction of the applied field, $[100]$.  
This non-collinear ferromagnetic 
structure allows the observed magnetic Bragg peaks at the $Q = 0$  
positions such as $2,0,0$.  
Interestingly the experimental magnetization did not develop smoothly, but 
in a series of steps.
Hysteresis was observed on cycling the field.

\begin{figure}
\includegraphics[width=0.46\textwidth]{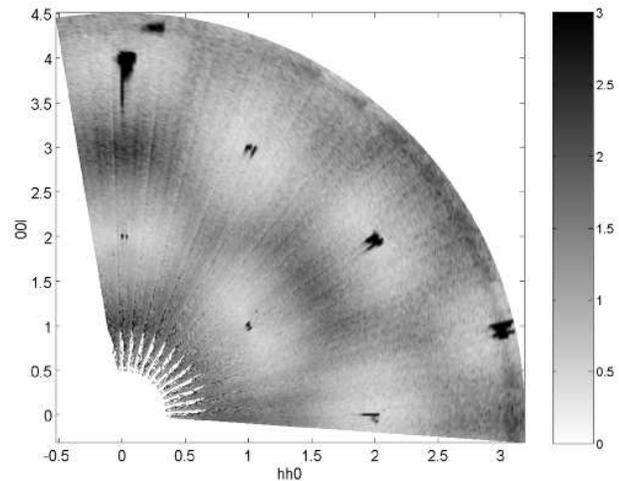}
\caption{Diffuse scattering in the $(hhl)$ plane.  In zero field at 270mK no magnetic Bragg peaks are observed.  All resolution limited intense features are of nuclear origin.}
\label{add}
\end{figure}

\begin{figure}
\includegraphics[width=0.46\textwidth]{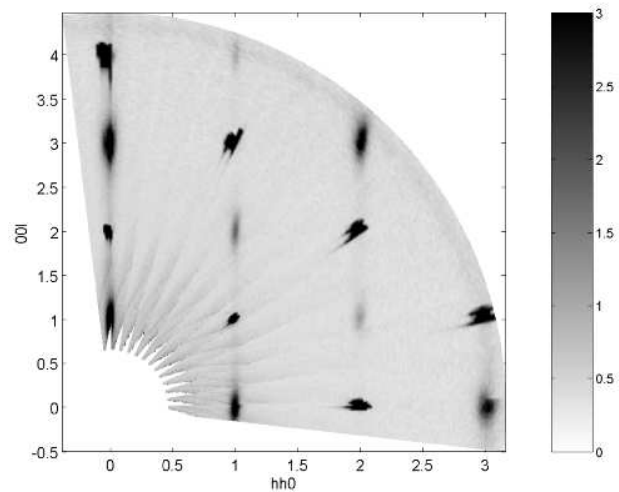}
\caption{Scattering in the $(hhl)$ plane with a field of 1.5T applied on 
$[1\bar{1}0]$ at $\approx 60$mK.  Magnetic Bragg peaks have appeared at 
positions such as 2,0,0 and the diffuse features observed in zero field at 
positions such as 0,0,3 have sharpened into features elongated on
$[00l]$.}
\label{tiger}
\end{figure}

\begin{figure}
\includegraphics[width=0.46\textwidth]{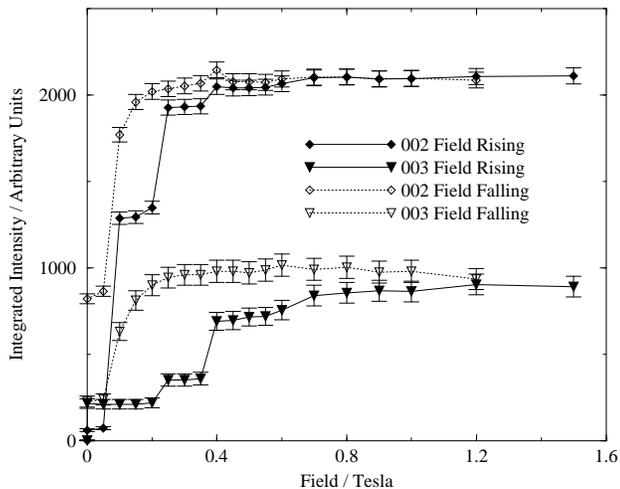}
\caption{Integrated intensity of the 0,0,2 magnetic Bragg peak and the 0,0,3 
diffuse feature (scaled$\times 3$ for this plot) as the field is scanned at $\approx 70$mK.}
\label{loop}
\end{figure}

The $[1\bar{1}0]$ direction is a hard direction of
magnetization \cite{Cornelius}. In zero field, the magnetic scattering was
observed to be diffuse, and characteristic of the disordered 
low temperature state of dipolar spin ice
\cite{newprl}, see Figure \ref{add}. 
Application of a small field ($\approx 0.1$ T) in this direction again caused
the appearance of the $Q = 0$ Bragg peaks; however, unlike the $[100]$ 
direction, the diffuse scattering features did not
disappear.  As the field was raised up to $1.5$T the  diffuse scattering 
sharpened around the $Q = X$ positions such as $0,0,1$ without becoming 
resolution limited. This is seen
clearly in figure \ref{tiger}. 
Similar features have been observed in the neutron scattering of  
Ho$_2$Ti$_2$O$_7$~\cite{prl1,dten}, adding strength to the idea that the true
ground state of these systems is a $Q = X$ structure that is dynamically
inhibited from being accessed on experimental timescales~\cite{Michel}. 
Again the magnetization versus
field curve was observed to have several sharp steps and plateaus 
(Figure \ref{loop}). 
 
\begin{figure}
\includegraphics[width=0.46\textwidth]{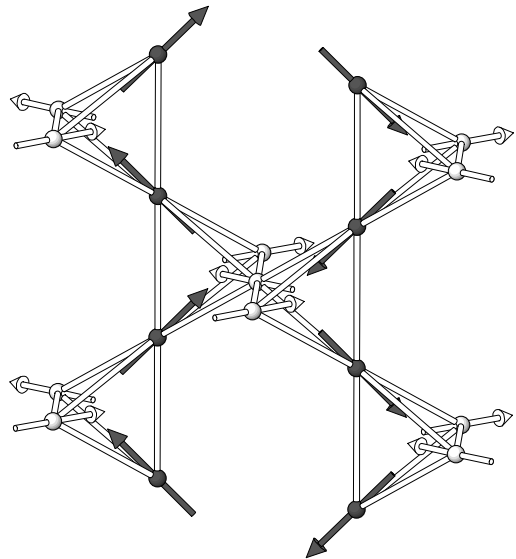}
\caption{A fragment of the pyrochlore lattice illustrating the $Q=X$ structure.  The applied field is projected into the plane of the page and pins the unshaded spins into ferromagnetic chains.  The shaded spins are not coupled to the field but are governed by the ``ice rules''.  The shaded spins may form two structures, either $Q=0$ with all shaded spin chains parallel or $Q=X$ with shaded spin chains antiparallel.  This degeneracy is not raised by the field.}
\label{xstr}
\end{figure}

The formation of the $Q = X$ structure is consistent with the spin ice 
rules \cite{prl1}.
In this field orientation, assuming perfect $\langle 111\rangle$
spins, only two of the spins of the tetrahedral basis have a component along
the field direction. These form ``in-out'' $[1\bar{1}0]$ 
chains parallel to the field
forcing the remaining two spins per tetrahedra into ``in-out'' 
$[110]$ chains
perpendicular to the field, as illustrated in Figure \ref{xstr}. The
perpendicular chains are not coupled by the spin ice rules which, in the
absence of any further neighbour coupling, would 
lead to two dimensional $[110]$ Bragg sheets
of scattering extended along $[00l]$ in the scattering plane. 
The diffuse features in the
experimental pattern are indeed extended along this direction, 
but the sharp build up of intensity around the $Q = X$ points
indicates a strong tendency to prefer $Q = X$ short range ordering of the
perpendicular rods (see Figure \ref{tiger}).  
Neglecting interference between the scattering from the 
two spin sets
(perpendicular and parallel to the field) one arrives at the conclusion that
the $Q = 0$ Bragg scattering arises from the parallel rods and the $Q = X$
diffuse scattering arises from the perpendicular rods. However, 
it may be a crude approximation to separate the scattering in this way. 

In conclusion, we have obtained accurate neutron scattering data for 
Dy$_2$Ti$_2$O$_7$ that is in qualitative agreement with theoretical
expectations for a spin ice material~\cite{prl1,newprl,jpcm,byron}. 
It is of interest that, even in 
a relatively strong field along $[1\bar{1}0]$, the system remains only
partially ordered. It is also noteworthy that the magnetic hysteresis loop
shows several steps and plateaus. An understanding of these effects awaits a
detailed study of the static and dynamic properties of the 
dipolar spin ice model~\cite{byron} in an applied magnetic field.

\section{Acknowledgements}
\label{ack}
We acknowledge the EPSRC for funding of beam time and studentships (TF 
and JDMC).

\end{document}